# On the Turbulence Beneath Finite Amplitude Water Waves

*Alexander V. Babanin* [1] *and Brian K. Haus* [2]

[1] *Swinburne University of Technology, Melbourne, Victoria 3122 Australia*
*ababanin@swin.edu.au*
[2] *University of Miami, Florida*

**Abstract**. The paper by Beya et al. (2012, hereinafter BPB) has a general title of "Turbulence beneath finite amplitude water waves", but is solely dedicated to discussing the experiment by Babanin and Haus (2009, hereinafter BH) who conducted measurements of wave-induced non-breaking turbulence by particle image velocimetry (PIV). The authors of BPB conclude that their observations contradict those of BH. Here we argue that the outcomes of BPB do not contradict BH. In addition, although the main conclusion of BPB is that there is no turbulence observed in their experiment, it actually is observed.

## 1. Introduction

The paper by Beya et al. (2012, hereinafter BPB) has a general title of "Turbulence beneath finite amplitude water waves", but is solely dedicated to discussing the experiment by Babanin and Haus (2009, hereinafter BH) who conducted measurements of wave-induced non-breaking turbulence by particle image velocimetry (PIV). The authors of BPB conclude that their observations contradict those of BH.

As we will argue below, contrary to what was published, the outcomes of BPB do not contradict BH. This has already been outlined by Babanin and Chalikov (2012) and Ghantous and Babanin (2014), but needs to be addressed explicitly since the BPB conclusions have been interpreted as evidence against the existence of wave-induced turbulence (e.g. D'Asaro. 2014).

As a reply to BPB, we will only briefly discuss the phenomenon of wave turbulence here. This is an important topic, which has been well described in the literature. Indeed, there is a long-standing tradition of approaching surface waves as an irrotational phenomenon which therefore cannot produce turbulence. This approach to wave dynamics, even for nonlinear waves, has been very successful throughout almost two centuries, with fundamental breakthroughs in the 60s (see e.g. Babanin et al. (2012) for a review). These breakthroughs made rotational solutions nearly forgotten. Rotational solutions, however, are well documented, and it is well established that vorticity in three-dimensional/non-homogeneous/random wave field leads to the presence of turbulence (e.g. Longuet-Higgins, 1953, 1960, Phillips, 1961, Kinsman, 1965). In an oceanographic context, the issue of rotationality is in fact secondary, as pre-existing three-dimensional turbulence is known to be unstable with respect to wave orbital motions, even if the latter is regarded as potential (Benilov et al., 1993, Benilov, 2012). Since the real ocean is always turbulent to some degree, there will be growth of this turbulence produced by surface waves and corresponding dissipation of the waves. Over the years, there have been various laboratory experiments, field observations and numerical simulations carried out to investigate such wave-induced turbulence and its influences (see Qiao et al. (2010), Babanin et al. (2012), Ghantous and Babanin (2014) and references wherein for more details).

## 2. Inaccuracies in BPB regarding the BH paper

There are several specific errors in BPB that must be addressed. These include their repeated statements that BH propose a wave-amplitude dependent critical wave number, their assertion that potential wave theories describe non-linear waves well and as such turbulence does not exist and their interpretation of their own experimental evidence, and other issues which separated into individual Sections below.

## 3. Critical Reynolds Number in BH

BPB make an emphasis on debunking the existence of a critical Wave Reynolds Number that they assert is proposed in BH. Their Abstract starts from statement: "Babanin and Haus (2009)... proposed a threshold wave parameter $a^2\omega/\nu = 3000$ for the spontaneous occurrence of turbulence beneath surface waves". Here, $a$ is wave amplitude, $\omega$ is radian frequency and $\nu$ is kinematic viscosity of water. This statement is then reiterated through the text and figure captions several times (e.g. page 1320 7th paragraph, page 1322 beginning section 3.1, caption Figure 1). This statement also restated as the final sentence of the conclusion of BPB.

Despite its centrality to BPB, this statement is completely innaccurate. BH did not propose a threshold for the spontaneous occurrence of turbulence. Moreover, BH explicitly state (page 2678): "Such intermittent turbulence does not reveal a threshold value of wave amplitude, below which it does not occur (i.e., the critical Reynolds number)". In BH Figure 2, which shows dependence of volumetric dissipation rate on the Wave Reynolds Number defined as the above, the value of 3000 is in the centre of the data, it is marked with asterisk and highlighted in the Figure caption as showing that no threshold value exists.

It is not possible to understand how the authors arrived at such an erroneous conclusion about BH, whose experiment they decided to reproduce thoroughly, but it is repeatedly apparent that BPB confuse intermittent turbulence with fully developed turbulence. They then proceed with a dye experiment, making references to the original experiment of Reynolds (1883), and do not seem to realize that the rapid dissolution of dye in Reynolds' tests indicated transition to the state of fully developed turbulence and not the presence of spontaneous turbulence. The critical Reynolds Number for a particular flow can vary, depending on initial and external flow conditions, but below this value turbulence can appear and be suppressed intermittently (see e.g. Turbulence chapter in the book on fundamentals of fluid mechanics by Landau and Lifshitz (1989)).

It is also likely that BPB confuse BH with an earlier paper by Babanin (2006). The latter is cited by BPB, but in a wrong context and with a wrong reference, as if the authors did not cite the paper as such. Babanin (2006) indeed inferred the critical Wave Reynolds Number of 3000, but that was done for the transition to the fully developed turbulence.

In summary, the statement that BH proposed a threshold Reynolds Number is fundamental to their entire manuscript and it is necessarily wrong.

**4. High-order wave solutions**

The second set of statements from the Abstract relates the non-existence of wave turbulence to the fact that nonlinear waves are well described by potential theory: "Many laboratory wave experiments were carried out in the early 1960s (e.g. Wiegel 1964). In those experiments, no evidence of turbulence was reported, and steep waves behaved as predicted by the high-order irrotational wave theories within the accuracy of the theories and experimental techniques at the time".

This argument does not seem to be justified. We first would like to comment that there hardly is a need to refer to the early 60s of the last century in this regard. By comparison with modern days, instrument precision and data recording capabilities were primitive then. And nowadays, such measurements are routine and comprehensive. For example, Babanin et al. (2010) measured waves with laser probes and compared with fully nonlinear numerical simulations rather than with high-order theories. They studied, among other issues, modulational instability of nonlinear wave trains. In 1964, such instability was yet to be discovered (Benjamin and Feir, 1967), but obviously experiments of Wiegel (1964) with nonlinear waves, which did not reveal such instability, do not disprove the fact of its existence. Particularly since as BPB admit the resolution of the visualization technique employed in Wiegel (1964) may not have been sufficient to detect the presence of turbulence.

The argument that predictions of high-order irrotational wave theories is well supported by laboratory observations of the shape and kinematics of Stokes waves is then elaborated in the text of BPB. Apparently, it is meant to imply that there is no room for rotational behavior of surface waves, i.e. used as an indirect evidence against wave turbulence. Such implication, however, disagree with the theory of wave motion in viscous fluids. The viscosity leads to rotational solutions with imaginary part (e.g. Kinsman, 1965). The imaginary part causes slow decay of Stokes solutions and do not contradict to the potential theory. Similarly, the mathematical theory of instability of three-dimensional pre-existing turbulence in presence of waves by Benilov (2012), does not impose any restrictions on the Stokes solution. In fact, Benilov instability works for potential waves.

In summary, observations of the nonlinear waves which confirm high-order Stokes theory can neither prove nor disprove the irrotational theories, and are unrelated to behavior of the water turbulence.

## 5. BPB experiment with dye flow visualisation

The main results of the paper are formulated towards the end of the Abstract as "flow visualization experiments for steep non-breaking waves using conventional dye experiments" which "showed no evidence of turbulent mixing" and "are in accord with the conventional understanding of wave behavior". In the Conclusions, the conventional understanding is clarified as "absence of turbulence beneath two-dimensional, freely propagating, unforced, non-breaking waves". It is also pointed out that "This is in contrast with the findings of BH2009 who reported turbulence levels comparable to field measurements under strong winds".

The last statement again highlights the confusion by BPB of intermittent turbulence in the BH laboratory tests and fully developed turbulence in the ocean. The problems of the BPB experiment and its interpretation, however, are more complex.

BH used a PIV system to measure the turbulent velocities directly. BPB made their conclusions on the absence of turbulence indirectly, by observing dye injected into the water. We must comment that overall, particularly by indirect inference, proving absence of a feature is a much more difficult task than proving its presence. For example, if a Mars probe did not find life, this fact does not prove that life does not exist on Mars. But if it did find the life, this is the proof of such existence.

Here, BPB never specified what is it that they expected to see in case of turbulence being present/absent, i.e. there is no guidance on criteria of their judgment. Since references are made to the Reynolds experiment, presumably they expected the dye to dissolve in the course of the observations.

We should stress that the dye will dissolve without any waves, currents, winds or turbulence. Therefore, in absence of direct measurements of turbulence, the judgment should have been made by means of comparison of the dissolution rates of dye and of molecular diffusivity. If the former is faster than the latter, then turbulence is present. For example, Dai et al. (2010) found that in presence of gently sloped non-breaking waves, vertical temperature stratification dissolves within 20 minutes. This is very slow, but still two orders of magnitude faster than in their absence, which was 20 hours. Such excessive dissipation rate can only be due to extra mixing, and the latter can only be due to presence of turbulence.

Another issue is the temporal scale of intermittency. For example, breaking of waves of regular steepness occurs at the scale of tens of wave periods (e.g. Babanin et al., 2010). If measurements are conducted over five wave periods, one can conclude that breaking does not happen, but this would apparently be an incorrect conclusion.

BPB conducted each observation over five wave periods. Since the turbulence dissolution rate of non-breaking waves is some 1000 wave periods (Dai et al., 2010), hardly anything of significance can be expected to happen within the five wave periods. Moreover, since the turbulence was intermittent in the BH setup, there may have indeed been no turbulence appearing over the duration of the tests in BPB.

In summary, results of flow visualization in BPB experiment are not in contradiction, but are in fact in agreement with BH who noted that "the Kolmogorov interval appeared from 0 to 3 times over a duration of 15 periods".

**6. Presence of turbulence in the BPB experiment**

The curious observation of the results demonstrated in BPB is that turbulence was in fact present in their experiments. This can be seen in their Figure 7, reproduced below.

In this Figure, the authors plot measurements on the dye line thickness (horizontal scale) away from the surface (vertical scale, surface is at the top). The three subplots correspond to different wave steepness *kH/2*, 0.17 (a, top), 0.21 (b, middle) and 0.24 (c, bottom). Here, *k* is wavenumber and *H* is wave height. Filled circles signify the initial conditions and hollow ones the thickness after 5 wave periods.

There is an apparent reduction of thickness of the line in cases a) and c), i.e. for the lowest and the highest steepness, but notably not for the medium steepness in panel b). The authors explain the reduction being a result of the "stretching of the dye line induced by the Stokes drift". In order to substantiate such explanation, they could have estimated such drift and quantified the expected stretching. Apparently, they compared their measurements with the fifth-order wave theory, and the Stokes drift is a second-order effect. In BPB, however, no attempt was made to do this, the authors just repeatedly postulate it.

In the meantime, the observed thinning of the line could not have been due to the Stokes drift, because of at least two reasons. First, the Stokes drift is proportional to the wave amplitude squared, and therefore in panel c) the reduction should have been twice as big as in panel a), at the same distance below the surface, whereas in fact it is smaller.

Most importantly, however, the Stokes drift could not have possibly be absent in panel b), which depicts waves of intermediate steepness, if it was present both for the higher steepness in panel c) and the lower steepness in panel a). Stokes drift is a deterministic effect, and in case of nonlinear waves it is always present, its relative strength grows as a function of steepness.

Therefore, the Figure shows us some intermittent effect on the dye line thickness: sometimes it is present, and sometimes is not. This can only be turbulence, since there was no breaking as the authors state.

Intensity of such turbulence is very strong by comparison with the measurements of Dai et al. (2010) where time scale of the dissolution was 1000 wave periods. Effectively, in both cases of its presence in BPB, the line thickness halved. Puzzlingly, however, strength of this turbulence is increasing away from the surface. Both the Stokes drift and intensity of the wave-induced turbulence must increase towards the surface. Therefore, the observed vertical behavior in panels a) and c) is either the turbulence produced at the bottom, or the picture is upside down (or perhaps the reflection of the image rather than the image is analyzed here). The latter option is more likely because Figures 3, 4, 8, 9 and 10 all show that thinning is larger towards the top rather than bottom.

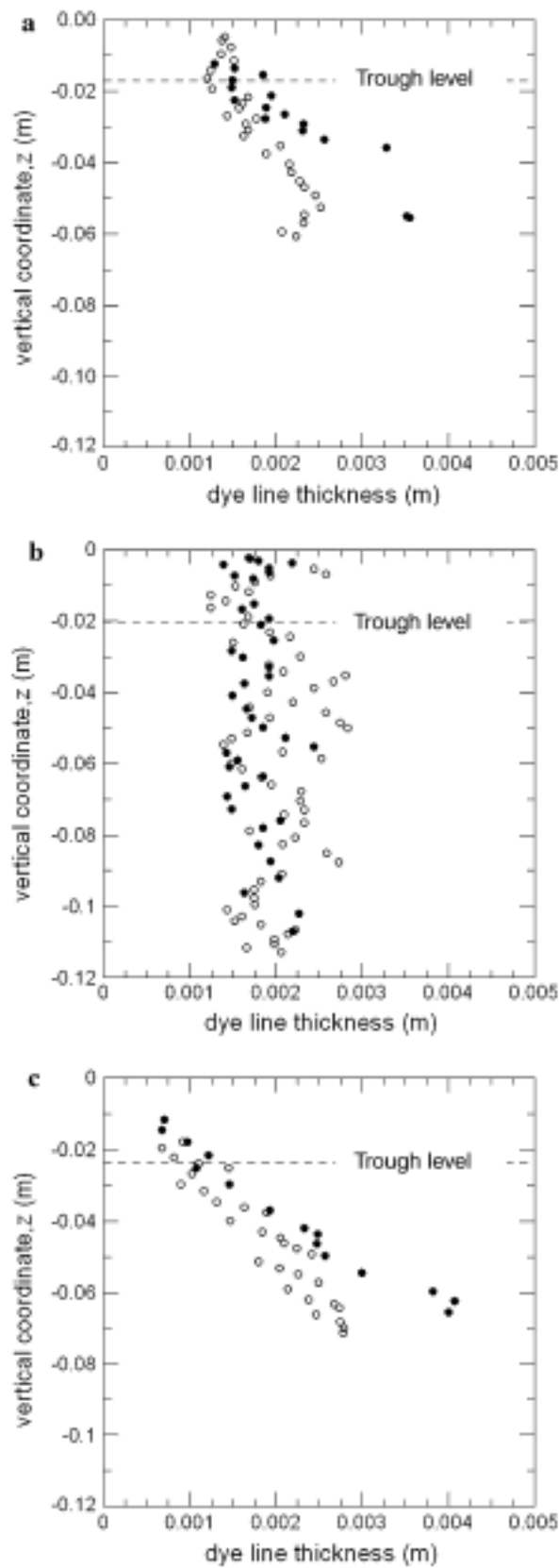

**Fig. 7** Vertical profile of dye line thickness at time ($t = 0$) (*solid circles*) and ($t = 5T$) (*hollow circles*). **a** Case 1, **b** Case 2, **c** Case 3. *Horizontal dashed lines* indicate estimated fifth-order theory trough levels. $z = 0$ is the still water level

In summary, turbulence was demonstrably present in the BPB experiment, very strong by the standards of the non-breaking wave turbulence and, unless this is the plotting mistake, increasing away from the water surface.

**7. ADV velocity measurements in BPB**

BPB also mention that velocity records were conducted by an Acoustic Doppler Velocimeter below wave troughs. They state that "The acquired time series and velocity spectra did not show any evidence of turbulent fluctuations". This again points to BPB's obvious misunderstanding of the nature of BH regarding the observation of " intermittent" turbulence.

In the paper, no details are provided which would allow to analyze this statement, but details are available in the Master Thesis of Beya (2011). Effectively, the BPB paper is and extract from this Thesis. According to Figure 5-11 of Chapter 5 in the Thesis, no Kolmogorov interval in velocity spectra is visible which fact is interpreted as absence of turbulence.

The highest frequency plotted in this Figure is 100 rad/s, and therefore, Beya (2011) was observing the spectral range mostly below the frequency range of the Kolmogorov interval in BH. Regardless of this, however, for intermittent turbulence the average spectrum derived from a single point time series cannot show such interval in principle. The -5/3 spectra in Figure 1 of BH are for those segments when such turbulence is present, and otherwise the respective segments show white noise in this frequency/wavenumber range. If the -5/3 segments and white-noise segments are averaged together they will display some unclear pattern depending on the relative weighting of the two groups. Still, spectra in Figure 5-11 of Beya (2011) do show slopes close to -5/3 rather than white noise beyond the 4$^{th}$ harmonic, which may be an outcome of the strong turbulence exhibiting itself in the Figure above as discussed.

In summary, the velocity spectra measurements in BPB could not cover the range of scales observed in BH and a single point time series will always obscure intermittent events when spectrally averaged. However, given these limitations, even Beya (2011) in a lower frequency spectral range may have shown the -5/3 interval.

**8. Discussion**

Towards the end of BPB, the authors list 5 items which they believe may have influenced the turbulence observations in BH. Some of them are trivial, others are reasonable, and we would like to address them before concluding.

Items 1 and 2 point out that the measurements should be taken clear from the front of the wave train and from the paddle. In the ASIST tank, where the BH experiment was conducted, the measurement site was 7 wavelengths from the paddle and the arrival time of the beach-reflected signal was estimated as 55 wave periods. Thus, the measurements were sufficiently far from the paddle (note that no background or advecting turbulence was observed), and the 15 periods of the BH records were kept well clear from both the original and the reflected fronts. The group modulation was undoubtedly present, as it was in BPB (see e.g. their Figure 6). Regardless of the paddle, this is natural behavior of the nonlinear groups which cannot be eliminated, and it may have contributed to the scatter and large confidence limits shown in BH. Item 3 refers to the PIV accuracy of 3 cm/s mentioned in BH. This was an apparent typo and certainly a regrettable error in BH, PIV systems are typically much better than this and the ASIST PIV accuracy was 3 mm/s or better, depending on the maximal time between images. This can be seen in BH by the 1.2 mm pixel size and the 4 ms maxium dwell time given. Also, even though this is not essential here, we should comment that in Eq. (1) of BPB, the similarity connection between right- and left-hand sides of the expression is fundamental, but the proportionality coefficient $A$ is not (see e.g. Landau and Lifshitz, 1987). For a different flow, or even for the same flow with different initial conditions, quantitative proportionality can be different, even by an order of magnitude. Items 4 and 5 are a comprehensive set of technical issues relating to turbulence measurements and spectral calculations if the turbulence is anisotropic and unsteady, waves are wind-forced, rotational, breaking and nonlinear. In the BH analysis, we used estimates of the volumetric dissipation rates $\varepsilon$ for the Kolmogorov cascade. This cascade establishes itself if there is energy input into the system at a scale, large with respect to the viscous dissipation scale. The Kolmogorov interval should exhibit itself as a -5/3 spectral slope, far enough from both scales. In our case, the large scale was 9 rad/m (1.5 Hz), Kolmogorov microscale $\sim$ 35000 rad/m, and the -5/3 interval was observed between 800 and 3000 rad/m. Other hypothetical events, such as micro-breaking which is difficult to observe are also mentioned by BPB, but obviously we cannot expect them to lead to the regular $\varepsilon \sim a^3$ behavior below wave troughs, particularly all the way to gently sloped waves.

## 9. Summary and Conclusions

BPB claims that their observations contradict to those by BH, but these claims are unsubstantiated and/or incorrect. The claims are addressed separately above, with a brief relevant summary at the end of each Section. Overall, we believe that BPB confused the BH paper with the earlier paper by Babanin (2006) which discussed fully developed turbulence. That paper established the critical Wave Reynolds Number and indeed conducted a dye experiment, but its results are not applicable to the intermittent turbulence in the BH experiment, as explicitly pointed out by BH. In addition, although the main conclusion of BPB is that there is no turbulence observed in their experiment, it actually is observed.

In this Summary, we would like to answer some more general issues raised by BPB. The authors admit that wave-induced turbulence is a missing dynamic of the upper ocean and, if reinstated, "would have significant and widespread implications for the entire air-sea interaction discipline". At the same time, from Abstract through Conclusions they advocate "conventional understanding of wave behavior" which is understood as irrotational and consequently non-turbulent.

We should explain here that the potential irrotational motion is only conventional as a solution of basic equations which assume the water to be non-viscous. If viscosity is introduced in the linear or nonlinear wave theories, then the conventional understanding is that the wave motion is rotational (e.g. Kinsman, 1965), and that the randomized or stretched vorticity is turbulent (e.g. Phillips, 1961). And once it exists, the vortices are unstable with respect to the wave orbital motion in planes perpendicular to these orbits (e.g. Benilov, 2012). This is mathematics which is not a subject of belief, and if not disproved mathematically is correct. Therefore, the existence of wave-induced turbulence is fundamental, even if conventionally disregarded in applications that rely on potential theory. The only relevant question in the context of fluid mechanics and physical oceanography is its relative strength and importance in various physical phenomena. We would also like to attract attention of the reader to laboratory experiments, numerical simulations and field observations of this turbulence other than BH (see e.g. Ghantous and Babanin (2014) for further references).

When conducting the BH experiment, we did not have any particular line of expectations and theory in mind, and the result of $\varepsilon \sim a^3$ was purely empirical. Since 2009, the BH results settled very well within further and independent developments of this topic. Babanin and Chalikov (2012) coupled a fully nonlinear wave model based on first principles with an LES turbulence model, and found volumetric dissipation rates in quantitative agreement with BH. Based on BH results, Babanin (2011) predicted swell dissipation rates, and Young et al. (2013) confirmed these rates in their satellite observations of swell propagation across the Southern Ocean. And of course completely independently, Bowden (1950) analytically suggested the same expression for the swell dissipation as Babanin (2011), including the BH proportionality $\varepsilon \sim a^3$. At the time of writing BH or Babanin (2011), Bowden's results were not known to us.